\documentstyle[amstex,preprint,aps]{revtex}
\begin{document}
\title{Experimental Detection of Entanglement with Polarized Photons}
\author{M. Barbieri, F. De Martini, G. Di Nepi, P. Mataloni}
\address{Dipartimento di Fisica and \\
Istituto Nazionale per la Fisica della Materia\\
Universit\`{a} di Roma ''La Sapienza'', Roma, 00185 - Italy}
\author{G. M. D'Ariano, C. Macchiavello}
\address{Quantum Optics and Information Group, \\
Istituto Nazionale per la Fisica della Materia, \\
Unit\`{a} di Pavia, \\
Dipartimento di Fisica ''A. Volta'', via Bassi 6,\\
27100 Pavia, Italy }
\maketitle

\begin{abstract}
We report on the first experimental realization of the entanglement witness
for polarization entangled photons. It represents a recently discovered
significant quantum information protocol which is based on few local
measurements. The present demonstration has been applied to the so-called
Werner states, a family of ''mixed'' quantum states that include both
entangled and non entangled states. These states have been generated by a
novel high brilliance source of entanglement which allows to continuously
tune the degree of mixedness.
\end{abstract}


One of the main issues of modern technology is the manipulation of {\it %
information}, its transmission, processing, storing, and computing, with an
increasingly high demand of speed, reliability and security. Quantum Physics
has recently opened the way to the realization of radically new
information-processing devices, with the possibility of guaranteed secure
cryptographic communications, and of huge speedups of some computational
tasks. In this respect quantum {\it entanglement} represents the basis of
the exponential parallelism of future quantum computers \cite{natureqc}, of
quantum teleportation \cite{telep1,telep2,telep3,telep4} and of some kinds
of cryptographic communications\cite{Ekert,Gisin}. In practical
realizations, however, entanglement is degraded by decoherence and
dissipation processes that result from unavoidable couplings with the
environment. Since entanglement is an expensive resource---it cannot be
distributed between distant parties by classical communication means---it
becomes crucial to be able to detect it efficiently, with the minimum number
of measurements. Several methods have been proposed to assess the presence
of entanglement for different types of quantum systems \cite
{susana,eh,exdet,Lew,rubin,our}. In particular, the socalled method of
''entanglement witness'' is a simple and efficient protocol that uses only a
few local measurements \cite{our}. In the present paper we report the first
experimental implementation of an entanglement witness for
polarization-entangled photons. The entangled photon state is generated by a
new general method of spontaneous parametric downconversion (SPDC) and the
entanglement is detected by only three independent quantum measurements.

Precisely, we implement experimentally the method of Ref. \cite{exdet} for a
pair of polarized photons that can be in any of the so-called Werner states 
\cite{werner}, a family of ''mixed''\ quantum states that include both
entangled and non entangled states. The key resource of our apparatus
consists of a new, universal high-brilliance source of bi-partite entangled
states spanning the Hilbert space $H_{1}\otimes H_{2}$, where $%
dim(H_{1})=dim(H_{1})=2$. These states can be generated as either ``pure''
or ''mixed'', with a complete control of the degree of mixedeness \cite
{ubsource,espwer}. In particular, here we generate the Werner states 
\begin{equation}
\rho _{W}=p|\Psi _{-}\rangle \langle \Psi _{-}|+\frac{1-p}{4}%
\mbox{$1\hspace{-1.0mm}  {\bf l}_4$}\;,  \label{werner}
\end{equation}
which are mixtures with probability $p\in \lbrack 0,1]$ of the maximally
chaotic state $\frac{1}{4}\mbox{$1 \hspace{-1.0mm}  {\bf l}$}_{4}$ ($%
\mbox{$1
\hspace{-1.0mm}  {\bf l}$}_{4}$ is the four dimensional identity operator)
and of the maximally entangled ''singlet'' state: 
\begin{equation}
|\Psi _{-}\rangle =%
%TCIMACRO{\tfrac{1}{\sqrt{2}}}%
%BeginExpansion
{\textstyle{1 \over \sqrt{2}}}%
%EndExpansion
(|HV\rangle -|VH\rangle ),  \label{Phi}
\end{equation}
where $|HV\rangle \equiv |H\rangle _{1}\otimes |V\rangle _{2}$ denotes the
state of the two photons, the symbols $H\ $and $V$ representing horizontal
and vertical linear polarizations, respectively. The method to establish
whether a state is entangled or not is based on the concept of entanglement
witness\cite{EW1,EW2}. According to this framework a state $\rho $ is
entangled if and only if there exists a Hermitian operator $W$---so called 
{\em entanglement witness}---which has positive expectation value $%
%TCIMACRO{\func{Tr}}%
%BeginExpansion
\mathop{\rm Tr}%
%EndExpansion
[W\rho _{sep}]\geq 0$ for all separable states $\rho _{sep}$, nevertheless
has negative expectation value $%
%TCIMACRO{\func{Tr}}%
%BeginExpansion
\mathop{\rm Tr}%
%EndExpansion
[W\rho ]<0$ on our state $\rho $. For pairs of qubits---the two-level
quantum systems of quantum information---in our case the polarized photons,
also the non positivity of the partial transposition of the state $\rho $
gives a necessary and sufficient criterion for entanglement \cite{peres,hor}%
. In this case simple ways to construct entanglement witnesses are known 
\cite{EW2}. The Werner states (\ref{Phi}) which are tested in our
experiment, are particularly appropriate for entanglement detection, because
they include both entangled ($p>1/3$) and separable ($p\leq 1/3$) states.

The detection method proposed in \cite{exdet} in the case of Werner states
expressed by (\ref{werner}) gives the following {\em entanglement witness}
operator 
\begin{equation}
\begin{split}
W& =%
%TCIMACRO{\tfrac{1}{2}}%
%BeginExpansion
{\textstyle{1 \over 2}}%
%EndExpansion
(|HH\rangle \langle HH|+|VV\rangle \langle VV|+|DD\rangle \langle DD| \\
& +|FF\rangle \langle FF|-|LR\rangle \langle LR|-|RL\rangle \langle RL|),
\end{split}
\label{witn}
\end{equation}
where $|D\rangle =\frac{1}{\sqrt{2}}(|H\rangle +|V\rangle )$ and $|F\rangle =%
\frac{1}{\sqrt{2}}(|H\rangle -|V\rangle )$ denote diagonally polarized
single photon states, while $|L\rangle =\frac{1}{\sqrt{2}}(|H\rangle
+i|V\rangle )$ and $|R\rangle =\frac{1}{\sqrt{2}}(|H\rangle -i|V\rangle )$
correspond to the left and right circular polarization states. The above
operator can be locally measured by choosing correlated measurement
settings, that allow detection of the linear, diagonal and circular
polarization for both photons. It represents the most efficient witness,
since it involves the minimum number of local measurements.

The experimental apparatus is shown in Fig. 1. A Type I, $.5mm$ thick, $%
\beta $-barium-borate (BBO) crystal is excited by a $V$-polarized cw $Ar^{+}$
laser beam ($\lambda _{p}=363.8nm$) with wavevector $-{\bf k}_{p}$, i.e.
directed towards the left in Fig.1. The two degenerate ($\lambda =727.6nm$)
SPDC\ photons have common $H$-polarization, and are emitted with {\it equal
probability, }over a corresponding pair of wavevectors belonging to the
surface of a cone with axis ${\bf k}_{p}$. The emitted radiation and the
laser beam are then back-reflected by a spherical mirror $M$ \ with
curvature radius $R=15cm$, highly reflecting$\ $both $\lambda $ and $\lambda
_{p}$, placed at a distance $d=R$ \ from the crystal. A zero-order $\lambda
/4$ waveplate placed between $M\ $and the BBO intercepts twice both
back-reflected $\lambda $ and $\lambda _{p}$ beams and then rotates by $\pi
/2$ the polarization of the back-reflected photons with wavelength $\lambda
\ $while leaving in its original polarization state the back-reflected pump
beam $\lambda _{p}\approx 2\lambda $. The back-reflected laser beam excites
an identical albeit distinct downconversion process with emission of a new
radiation cone directed towards the right in Fig.1 with axis ${\bf k}_{p}$.
In this way each pair originally generated towards the left in Fig. 1 is
made, by optical back-reflection and a unitary polarization-flipping
transformation, ''{\it in principle} {\it indistinguishable''} with another
pair originally generated towards the right and carrying the state $\left|
HH\right\rangle $. The state of the overall radiation, resulting from the
two overlapping indistinguishable cones, is then expressed by the pure
entangled Bell-state: 
\begin{equation}
|\Phi \rangle =%
%TCIMACRO{\tfrac{1}{\sqrt{2}}}%
%BeginExpansion
{\textstyle{1 \over \sqrt{2}}}%
%EndExpansion
\left( |HH\rangle +e^{i\phi }|VV\rangle \right)  \label{phi1}
\end{equation}
with phase $(0\leq \phi \leq \pi )$\ reliably controlled by micrometric
displacements $\Delta d\ $of $M$\ along ${\bf k}_{p}$. A positive lens
transforms the overall emission \ {\it conical }distribution\ into a {\it %
cylindrical }one \ with axis ${\bf k}_{p}$, whose transverse circular
section identified the ''{\it Entanglement-ring''}. Each couple of \ points
symmetrically opposed through the center of the ring are then correlated by
quantum entanglement. An annular mask with diameter $D=1.5cm$ and width $%
\delta =.07cm$ provides an accurate spatial selection of the ring. This is
divided in two equal portions along a vertical axis by a prism-like
two-mirror system and detected by two independent silicon-avalanche
photodiodes, mod. SPCM-AQR14 at sites $A$ and $B$. Typically, two equal
interference filters,\ placed in front of the $A$ and $B$detectors, with
bandwidth $\Delta \lambda =6nm$, determine the {\it coherence-time }of the
emitted photons:\ $\tau _{coh}${\it \ }$\approx 140$ f$\sec $. More than $%
4000$ coincidences per second are detected for a pump power $P_{p}$ $\simeq
100mW$.

Werner states (\ref{werner}) are generated by selecting a convenient {\it %
patchwork } technique which implies the following steps \cite{espwer}: [{\it %
i}] Making reference to the original {\it source-state} expressed by Eq. (%
\ref{phi1}), the {\it singlet }state $|\Psi _{-}\rangle $ is easily obtained
by inserting a zero-order $\lambda /2$ waveplate in front of detector $B$. [%
{\it ii}]{\bf \ }A anti-reflection coated glass-plate ${\it G}$, $200\mu m\ $%
thick, inserted between $M$ and BBO with a variable transverse position $%
\Delta x$, introduces a decohering fixed time-delay $\Delta t>\tau _{coh}\ $%
that spoils the{\it \ \ indistinguishability} of the {\it intercepted
portions} of the overlapping {\it quantum-interfering} radiation cones:
Fig.1, inset. As a consequence, {\it all nondiagonal} \ elements of $\rho
_{W}$ given by the surface sectors ${\bf B}+{\bf C}$ of the
Entanglement-ring, the ones optically intercepted by $G$, are set to {\it %
zero }while the non intercepted sector ${\bf A}$ expresses the singlet
contribution to $\rho _{W}$. [{\it iii}] A $\lambda /2$ waveplate is
inserted in the semi-cylindrical photon distribution reflected by the
beam-splitting prism towards the detector $A$. Its position is carefully
adjusted in order to intercepts {\it half }of the ${\bf B}+{\bf C}\ $sector,
i.e. by making ${\bf B}={\bf C}$. Note that only {\it half }of the ring
which is represented in Fig.1 inset needs to be intercepted by the optical
plates, in virtue of the EPR nonlocality. In summary, the sector ${\bf A}$
contributes to $\rho _{W}$ with probability $p$ with the pure state $|\Psi
_{-}\rangle \left\langle \Psi _{-}\right| $, the sector ${\bf B}+{\bf C}=2%
{\bf B}$ with probability $1-p$ with the mixture: 
\begin{equation}
%TCIMACRO{\tfrac{1}{4}}%
%BeginExpansion
{\textstyle{1 \over 4}}%
%EndExpansion
\left[ \left| HV\right\rangle \left\langle HV\right| +\left| VH\right\rangle
\left\langle VH\right| +\left| HH\right\rangle \left\langle HH\right|
+\left| VV\right\rangle \left\langle VV\right| \right]  \label{mixt}
\end{equation}
and the probability $p$ can be easily varied over its full range of values,
going from $p=0$ ($\rho _{W}\equiv \frac{1}{4}%
\mbox{$1\hspace{-1.0mm}  {\bf
l}_4$}$) to $p=1$ ($\rho _{W}\equiv |\Psi _{-}\rangle \left\langle \Psi
_{-}\right| $).

In order to detect whether the produced Werner state is entangled or not,
the expectation value of the witness operator (\ref{witn}) has been
measured, by performing local correlated measurements on each arm of the
linear, diagonal and circular polarization. The polarization of the detected
photon is selected by means of a sequence of $\frac{\lambda }{4}$ and $\frac{%
\lambda }{2}$ waveplates, and a polarization beam splitter. Several
different values of the singlet weight $p$ have been tested in the
experiment. Each measurement corresponding to the different probabilities of
Eq. (\ref{witn}) which contribute to evaluate the entanglement witness has
been performed with an average of $30\sec $. The relative results are shown
in Fig. 2, where the experimental expectation value of $W$ is plotted as a
function of $p$, with the corresponding error bars. The agreement of the
experimental results with the theoretical prediction $%
%TCIMACRO{\func{Tr}}%
%BeginExpansion
\mathop{\rm Tr}%
%EndExpansion
[W\rho _{W}]=(1-3p)/4$ appears very good. The experimental results verify
the transition between separable and entangled Werner states, occurring at $%
p=1/3$, \ $%
%TCIMACRO{\func{Tr}}%
%BeginExpansion
\mathop{\rm Tr}%
%EndExpansion
[W\rho _{W}]=0$, as expected.

This work has been supported by the FET European Networks on Quantum
Information and Communication Contract IST-2000-29681:ATESIT and Contract
IST-2002-38877: QUPRODIS, MIUR 2002-Cofinanziamento, and by PRA-INFM\ 2002
CLON.

\centerline{\bf Figure Captions}

\vskip 8mm

\parindent=0pt

\parskip=3mm

FIGURE 1. Scheme of the experimental apparatus. The polarization entangled
photons are generated by spontaneous parametric down conversion in a
nonlinear BBO crystal, and are detected by two silicon-avalanche photodiodes
preceded by polarization analyzers. In this experiment we exploit a novel
geometry for collecting down-converted radiation, which results in a very
bright source of entangled photons, with also the advantage of complete
freedom in choosing the two-photon state. Inset: partition of the
Entanglement-ring into the spatial contributions of the emitted pair
distribution to an output Werner-state.

FIGURE 2. Experimental results of entanglement detection for Werner states.
These states range from the pure singlet at weight $p=1$ to the totally
chaotic state at $p=0$ (see \ref{Phi}, with the transition between entangled
and separable states at $p=1/3$. The entanglement is detected by the witness
observable given in \ref{witn}. The straight line corresponds to the
theoretical prediction. The experimental results verify the transition
between separable and entangled Werner states, occurring at zero-witness at $%
p=1/3$ (dotted line).


\begin{references}
\bibitem{natureqc}  Ekert, A., Jozsa, R., {\it Rev. Mod. Phys.} {\bf 68},
733-753 (1996); Knill, E., Laflamme, R., Milburn, G.J., {\it Nature} {\bf 409%
}, 46-52 (2001).

\bibitem{telep1}  Bennett, C.H. {\em et al.}, {\it Phys. Rev. Lett}. {\bf 70}%
, 1895-1898 (1993).

\bibitem{telep2}  Bouwmeester, D. {\em et al.}, {\it Nature} {\bf 390},
575-579 (1997).

\bibitem{telep3}  Boschi, D. {\em et al.}, {\it Phys.\ Rev.\ Lett.}\ {\bf 80}%
, 1121-1125 (1998).

\bibitem{telep4}  Marcikic, I. {\em et al.}, {\it Nature} {\bf 421}, 509-513
(2003).

\bibitem{Ekert}  Ekert, A., {\it Nature} {\bf 358}, 14-15 (1992).

\bibitem{Gisin}  Gisin, N. {\em et al.}, {\it Rev. Mod. Phys.} {\bf 74},
145-195 (2002).

\bibitem{susana}  Sancho, J.M.G., Huelga, S.F., {\it Phys. Rev. A} {\bf 61},
042303 (2000).

\bibitem{eh}  Horodecki, P., Ekert, A., {\it Phys. Rev. Lett}. {\bf 89},
127902 (2002).

\bibitem{exdet}  G\"{u}hne, O. {\em et al.}, {\it Phys. Rev. A} {\bf 66},
062305 (2002)

\bibitem{Lew}  G\"{u}hne, O. {\em et al.} \ e-print quant-ph/0210134.

\bibitem{rubin}  Pittenger, A.O., Rubin, M.H., {\it Phys. Rev. A} {\bf 67},
012327 (2003).

\bibitem{our}  D'Ariano, G.M., Macchiavello, C., Paris, M.G.A., {\it Phys.
Rev. A} {\bf 87} 042310 (2003).

\bibitem{werner}  Werner, R., {\it Phys. Rev. A} {\bf 40}, 4277-4281 (1989).

\bibitem{ubsource}  Giorgi G., Di Nepi G., Mataloni P., De Martini F. A, 
{\it Laser Physics} {\bf 13}, 350 (2003).

\bibitem{espwer}  Barbieri, M., De Martini, F., Di Nepi, G., Mataloni, P.,
e-print quant-ph/0303018.

\bibitem{EW1}  Terhal, B., {\it Lin. Alg. Appl.} {\bf 323}, 61-73 (2001).

\bibitem{EW2}  Lewenstein, M. {\it et al.}, {\it Phys. Rev. A} {\bf 62},
052310 (2000); Lewenstein, M. {\it et al., ibid.} {\bf 63}, 044304 (2001);
Bru\ss \thinspace\ D. {\it et al.}, {\it J. Mod. Opt.} {\bf 49}, 1399-1418
(2002).

\bibitem{peres}  Peres, A., {\it Phys. Rev. Lett.} {\bf 77}, 1413-1415
(1996).

\bibitem{hor}  Horodecki, M., Horodecki, P., Horodecki, R., {\it Phys. Lett.
A} {\bf 223}, 1-8 (1996).
\end{references}
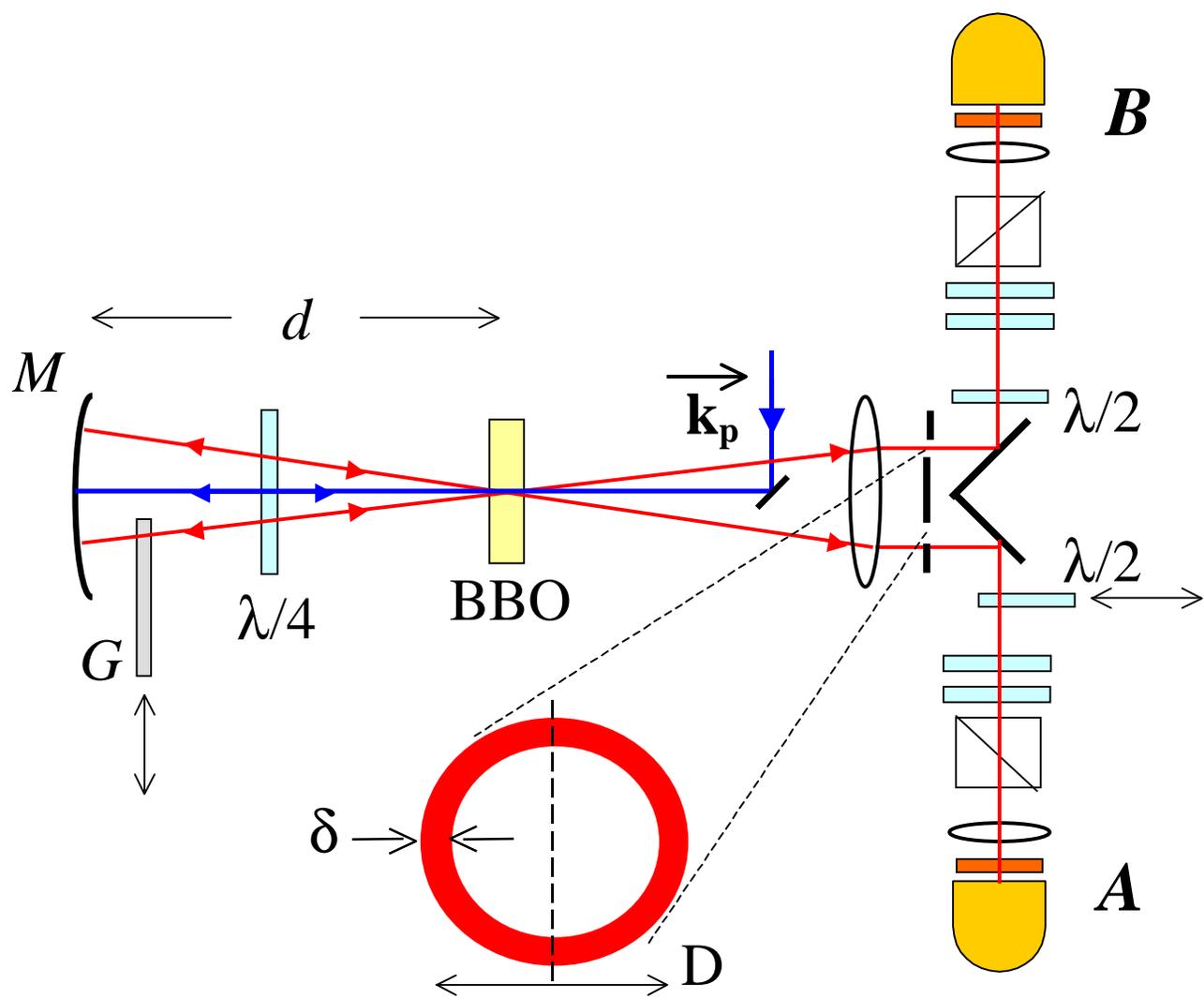
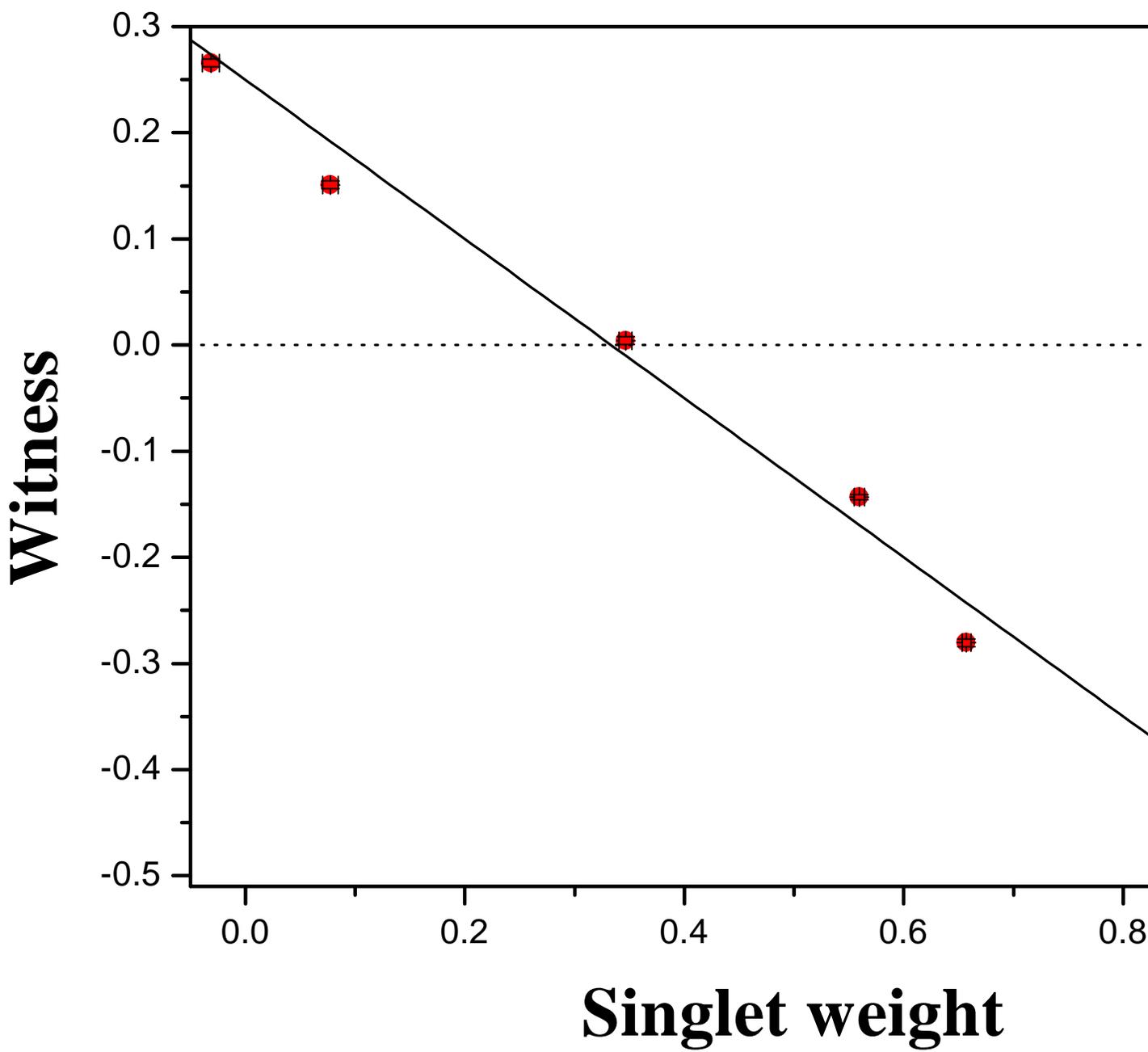
\end{document}